\documentclass[conference]{IEEEtran}
\IEEEoverridecommandlockouts
\usepackage{amsmath,amsfonts}
\usepackage{algorithm}
\usepackage{array}
\usepackage[caption=false,font=normalsize,labelfont=sf,textfont=sf]{subfig}
\usepackage{textcomp}
\usepackage{stfloats}
\usepackage{url}
\usepackage{verbatim}
\usepackage{graphicx}
\usepackage{cite}
\usepackage{listings}
\usepackage{algpseudocode}
\usepackage{booktabs}
\usepackage{ragged2e}
\usepackage{pgfplots}
\usepackage{tikz}
\usetikzlibrary{positioning}
\pgfplotsset{width=10cm,compat=1.17}
\algnewcommand\algorithmicnot{\textbf{not}}
\algdef{SE}[IF]{IfNot}{EndIf}[1]{\algorithmicif\ \algorithmicnot\ #1\ \algorithmicthen}{\algorithmicend\ \algorithmicif}%
\usepackage[section]{placeins}
\usepackage{graphicx}
\usepackage[justification=centering]{caption}
\def\BibTeX{{\rm B\kern-.05em{\sc i\kern-.025em b}\kern-.08em
    T\kern-.1667em\lower.7ex\hbox{E}\kern-.125emX}}
\begin{document}

\title{Comparative Study of Virtual Machines and Containers for DevOps Developers\\
\thanks{*Corresponding author; ORCiD: 0000-0002-3757-6463}
\thanks{S. Deochake, S. Maheshwari, R. De, and A. Grover performed this research at Rutgers University, NJ, USA.}
}

\author{\IEEEauthorblockN{Saurabh Deochake*}
\IEEEauthorblockA{
\textit{Rutgers University}\\
New Brunswick, NJ \\
saurabh.deochake@rutgers.edu
}
\and
\IEEEauthorblockN{Sumit Maheshwari}
\IEEEauthorblockA{
\textit{Rutgers University}\\
New Brunswick, NJ \\
sumit.maheshwari@rutgers.edu}
\and
\IEEEauthorblockN{Ridip De}
\IEEEauthorblockA{
\textit{Rutgers University}\\
New Brunswick, NJ \\
ridip.de@rutgers.edu}
\and
\IEEEauthorblockN{Anish Grover}
\IEEEauthorblockA{
\textit{Rutgers University}\\
New Brunswick, NJ \\
anish.grover@rutgers.edu}
}

\maketitle

\begin{abstract}
This paper presents a comparative study of virtual machines (VMs) and containers for DevOps developers. The study explores the benefits and drawbacks of each technology in terms of their functionality, performance, security, and resource utilization. The paper examines the underlying architecture of VMs and containers, and how they differ from each other. The study includes a series of experiments that compare the performance and resource utilization of VMs and containers in different scenarios. The experiments evaluate factors such as startup time, memory usage, disk I/O, network latency, scalability, and administrative overhead. Finally, the paper provides recommendations for DevOps developers on which technology to choose based on their specific requirements and use cases. Overall, the study aims to provide a comprehensive understanding of the strengths and limitations of VMs and containers, helping developers to make informed decisions when choosing between them.
\end{abstract}


\section{Introduction}\label{section:introduction}
\IEEEPARstart{V}{irtualization} is a crucial technique driving several research areas in today's world. Virtualization has become a fundamental technique in modern computing, enabling the creation of multiple isolated virtual environments on a single physical host. Virtualization has played a crucial role in advancing several research areas, including cloud computing, high-performance computing, and scientific computing. In the past, private cloud computing environments were set up using physical servers in a data center, requiring expensive hardware and maintenance costs. However, with the advent of virtualization, cloud infrastructure costs have significantly decreased. By creating a virtual version of machine hardware, storage devices, and network devices using emulators called Hypervisors, virtualization has become an industry-standard deployment method for production software. However, as cloud computing matured, it became apparent that deploying virtual machines caused overhead, which adds up significantly when hundreds of virtual machines are deployed. Each virtual machine requires its own operating system and software stack, leading to duplication of resources and administrative complexity. To address this issue, virtual environments as alternatives to full-blown virtual machines were implemented.

One such virtual environment is Linux Containers (LXC) \cite{bib1}, which allows for lightweight virtualization compared to virtual machines. With LXC, a single host can run multiple isolated containers, each with its own file system, network interfaces, and process space. Containers share the same host operating system, avoiding duplication of resources and management overhead. Another example of a virtual environment is FreeBSD Jails, which allows for the creation of secure and isolated environments on a FreeBSD system. Jails provide a lightweight alternative to virtual machines, with lower overhead and faster startup time.

The concept of containers revolutionized the operations sector of software development, enabling DevOps developers to deploy applications quickly and efficiently \cite{bib2}. Containers provide a more lightweight and efficient alternative to virtual machines, allowing for faster startup times, higher density, and lower overhead. By eliminating infrastructure costs and overhead associated with virtual machines, containers have become the de-facto deployment method for modern cloud-native applications.

In this paper, we will discuss the trade-offs of using virtual machines and containers for deployment of software in the DevOps perspective. We will also study the impact of using Docker containers to deploy software to clients and benchmark the performance of virtual machines and containers. It is often claimed that container deployment times are significantly faster than virtual machines, reducing deployment times from days to minutes. Our paper will explore the reasoning and meaning behind this statement.

The paper is structured as follows: Section \ref{section:background} revisits the conceptual context on virtualization and containerization, section \ref{section:vm-or-container} explores scenarios for choosing between virtual machines or containers, section \ref{section:experiment} showcases our experiment, section \ref{section:results} discusses the results of the experiment, section \ref{section:analysis} analyzes shortcomings of container technology, and finally, we conclude this paper in section \ref{section:conclusion}.

\section{Background}\label{section:background}
This section aims to provide an overview of the basics of virtual machines, containers, and DevOps as an important part of software development life cycle on cloud computing platforms. 

\subsection{Cloud Computing}
Cloud computing and virtualization are often confused, but while these technologies share some similarities, they are not exactly the same. Virtualization is a technique that separates physical architecture into various dedicated resources, maximizing the efficiency of hardware by creating virtual versions of machine hardware, storage devices, and network devices. On the other hand, cloud computing is a service that results from the manipulation of hardware carried out by virtualization \cite{bib3}. Today, these two technologies are often combined to achieve the best results, using the virtualized environment over the cloud. Virtualization enables multiple operating systems and applications to run on the same server, leading to increased efficiency and reduced costs.

The evolution of cloud computing has made it even easier for small-scale businesses to benefit from software as a service (SaaS) applications, which allow businesses to pay only for the resources they use \cite{bib4}. While both cloud computing and virtualization offer benefits, we consider cloud computing as an extension of virtualization using infrastructure as a service (IaaS).

\subsection{Virtual Machines}
A virtual machine (VM) is a software implementation of a computer system that can run applications and operating systems in an isolated environment. A hypervisor, a specialized software, creates a virtualized layer over the physical hardware, which is then used by virtual machines to emulate the necessary hardware resources \cite{bib5}. The hypervisor manages and allocates the physical resources such as CPU, memory, and storage, to the virtual machines running on top of it. This enables multiple virtual machines to share the resources of a single physical machine, thereby reducing the cost of additional hardware. The consolidation of hardware resources leads to a reduction in power and cooling demands, which in turn reduces the management efforts in managing resources. Industry leaders such as VMware, Oracle, and Microsoft have developed and popularized VM technology.

\begin{figure}[htbp]
  \centering
  \includegraphics[width=\columnwidth]{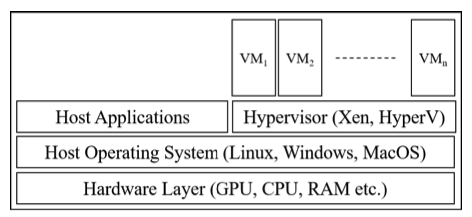}
  \caption{Virtual Machine Architecture}
  \label{fig:vm}
\end{figure}

\subsection{Containers}
Containers are platforms that enable developers to quickly develop and deploy applications, without worrying about the underlying infrastructure. They allow applications to run in a loosely isolated environment, making it possible to run many containers simultaneously on a single host. One of the key benefits of containers over virtual machines is that they don't require a hypervisor, which reduces overhead and enables faster deployment \cite{bib9}. Containers are powered by a kernel feature called "chroot", which provides the ability to change the root directory of a process, and restricts access to files outside the designated directory \cite{bib6}. Docker is a well-known container platform that provides tools for managing the lifecycle of containers.

\begin{figure}[htbp]
  \centering
  \includegraphics[width=\columnwidth]{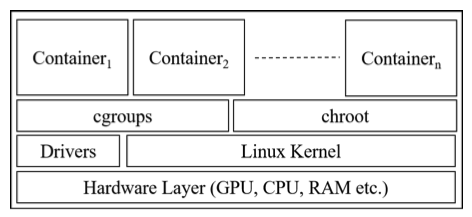}
  \caption{Virtual Machine Architecture}
  \label{fig:container}
\end{figure}

\subsection{DevOps}
DevOps is a software development methodology that aims to bridge the gap between software development and operations \cite{bib7}. It facilitates continuous integration and deployment of software by providing a collaborative and automated approach to software development. This methodology involves various stages of software development like building, testing, and deployment of software components. During the deployment phase of software development process, DevOps professionals can choose between using a virtual machine or a container. Virtual machines offer high adaptability and provide complete isolation of the resources, making them more secure and reliable. However, they are heavier and consume more resources than containers. Containers, on the other hand, primarily focus on applications and their dependencies. They run on a single kernel and can be easily deployed and maintained. Additionally, multiple containers can be deployed on a single host or virtual machine, but there is a higher risk of all the containers going down if the host goes down. Therefore, choosing between virtual machines and containers depends on the specific needs of the project, such as the level of security and reliability required, and the available resources.

\section{Choosing Between VMs and Containers}\label{section:vm-or-container}
Virtual machines and containers provide different ways to virtualize resources for running applications. In the case of a virtual machine, the infrastructure layer, known as the hypervisor, partitions the server below the operating system, creating true virtual machines that only share hardware resources. On the other hand, container virtualization is done at the operating system level, where some middleware is shared. Virtual machines offer higher flexibility since applications can run on a bare metal server, allowing users to select their own operating system and middleware. However, with containers, users have to choose a common operating system and middleware for their application. If you need a full platform to run multiple services, virtual machines are a better option. However, if you want to deploy a scalable service on a distributed platform, containers should be used. In this section, we have presented the following factors that determine which technology should be used and when.

\subsection{Operating System Requirements}
As a DevOps engineer, the person responsible for managing the development and deployment of software, one can select an operating system and middleware as per their choice. This choice of the operating system could be independent of other VMs or containers running on the same server. However, this is not the case with containers. The engineer would have to provide a "common" operating system and middleware elements when running the applications. This is because each container would use the core server platform and share it with other containers.

Virtual Machines offer a high level of flexibility, mainly due to the fact that the applications running on the guest environment are similar to a bare-metal server. On the other hand, containers are more focused on the application and its dependencies, and their deployment and management are easier compared to virtual machines.

\subsection{Fault Tolerance}
As a DevOps engineer, it's important to consider fault tolerance when choosing between virtual machines and containers. Virtual machines offer a demarcation between the operating system and physical hardware, creating an instance with its own set of services, such as virtualized adapters for network, storage units, CPU, and a replicated operating system. This means that applications running on a virtual machine are oblivious to the system and hardware resources, resulting in a restricted control to the system resources. As a result, if one virtual machine fails, it's less likely to affect other running virtual machines.

On the other hand, containers share kernel resources and application libraries, and thus, applications running on containers are system-aware with no hardware isolation. As a result, the application can take control of the system resources, which increases the possibility of one container failing and affecting other running containers. Therefore, when it comes to fault tolerance, virtual machines are a better option as they offer more hardware isolation and restrict control over the system resources, making it less likely for the failure of one virtual machine to affect other running virtual machines.

\subsection{Scope of the Application}
When choosing between virtual machines and containers, it's important to consider the nature of the application. Containers are more flexible and can run in any environment, regardless of the infrastructure, making them well-suited for web applications and small databases. On the other hand, virtual machines partition the server below the operating system and share the hardware, providing better isolation and security. As a result, they are more appropriate for embedded systems and infrastructure applications that require a higher level of fault tolerance and stability. On the other hand, containers provide the ability to package and run an application in a loosely isolated environment, making them very portable and easy to deploy.

\subsection{Scale of the Application}
Running applications on containers is the best option if there is a need to maximize the number of applications on minimum servers. Deploying multiple instances of a single application in multiple containers is less troublesome compared to Virtual Machines. Containers are highly useful in microservices, where each container runs a single service, and they can be scaled quickly. Additionally, deploying multiple web servers in containers requires significantly less hardware than on virtual machines. However, for running large applications or databases that require multiple machines, virtual machines would be a better choice.

\subsection{Service Model}
When deciding between containers and virtual machines, it's important to consider the service model used to deploy and manage environments. Virtual machines are typically deployed using tools like VMware, which handle the creation and migration of the VM to other environments. However, these tools may not be as useful when it comes to application management with DevOps, as they tend to be more focused on infrastructure management rather than software development and deployment. Therefore, containers have become increasingly popular in DevOps environments because they allow for greater flexibility and faster deployment of applications, making it easier to manage and maintain software across different environments.

\subsection{Organizational Constraints}
When it comes to selecting between virtual machines and containers, organizational constraints can also play a significant role. If an organization has the requirement to run different applications on a variety of software platforms, virtual machines may be the better option. This is because containers require a standardized hosting platform, which can make them more difficult to use in a heterogeneous environment. On the other hand, if a software application is dependent on a specific version of the operating system and there is a need to run multiple instances of the application, containers can be a better choice. This is because deploying multiple instances of an application in containers is generally easier and can result in more efficient resource utilization. Ultimately, the decision between virtual machines and containers will depend on the specific organizational requirements and constraints involved \cite{bib8}.

\section{Experiment}\label{section:experiment}
This section discusses the steps that we took to measure the effectiveness of Docker containers over virtual machines for deploying software. In addition to measuring the time required for each virtualization technique to start operating, we also measured network performance, disk I/O, CPU usage, and other administrative functionalities such as checkpoint and restore. For our test environment, we deployed a simple Apache web server that serves a website performing heavy mathematical functions like a naive implementation of the Fibonacci number series. We used both Docker containers and virtual machines to deploy the web server, which was backed by a PostgreSQL database server providing data to the Apache web server. Each test is performed four times to avoid any skewed results. Our test environment comprised of the following components.

\begin{itemize}
  \item Virtual Box or QEMU KVM as a virtual machine
  \item Docker containers
  \item Kubernetes to orchestrate the containers
  \item System benchmarking tools
  \item Apache web server
  \item PostgreSQL database server
\end{itemize}

\subsection{System Specification}
\subsubsection{Virtual Machine}
To study our web application using a virtual machine, we are using the Ubuntu Linux distribution and testing our Apache-based web service. The system specifications being utilized for this purpose are as follows:

\begin{itemize}
  \item Operating system: Ubuntu 16:04
  \item Linux kernel: v4.10
  \item Hypervisor: Oracle VM Type-2 Hypervisor
  \item Apache web server: httpd v2.24
  \item PostgreSQL: v9.6
\end{itemize}

\begin{figure}[htbp]
  \centering
  \includegraphics[width=\columnwidth]{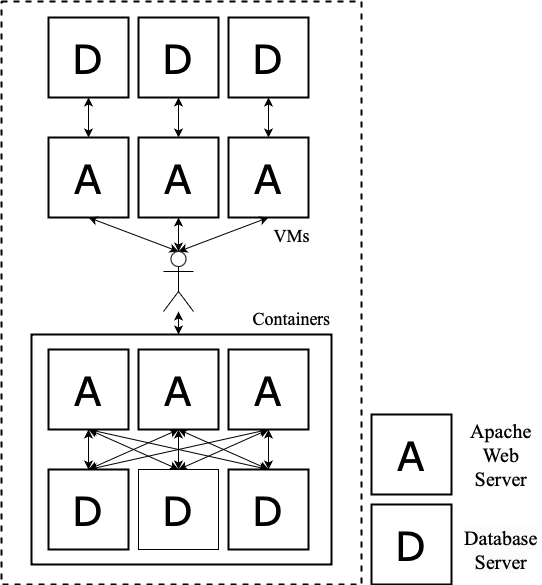}
  \caption{Experiment Setup}
  \label{fig:arch}
\end{figure}

\subsubsection{Containers}
After deploying the web application on virtual machines, we proceeded to test the same deployment on Docker containers. The following are the container specifications that we used to study our web applications:
\begin{itemize}
  \item Docker: v17.09-ce
  \item Operating system image: ubuntu:latest
  \item Linux kernel: v4.10
  \item Apache web server: httpd v2.24
  \item PostgreSQL: v9.6
\end{itemize}

To ensure accurate results, it is important to use identical versions of Apache web server, PostgreSQL server, and Linux Kernel to avoid any discrepancies caused by differences in version numbers of Kernel modules. We evaluate virtual machines and Docker containers based on DevOps considerations such as the time it takes to set up the environment, the time it takes to start and stop services, the ease of operation including continuous integration and deployment, and the ease of scaling the application. We also compare the resource management requirements for each approach, for both virtual machines and containers. Ultimately, we will determine whether containers are the best approach or whether virtual machines offer more benefits for DevOps developers.

\begin{verbatim}
$ docker pull debian:jessie
$ docker images
$ docker run - ti 25fc9eb3417
$ docker run - ti -d 25fc9eb3417
\end{verbatim}

\section{Results}{\label{section:results}}
This section presents the findings of the study comparing the effectiveness of Docker containers and virtual machines in deploying a software. The section provides a detailed analysis of various metrics such as time to set up the environment, time required to start and stop services, ease of operation, scalability, and resource management. The section presents a comparison of the performance of virtual machines and Docker containers on these metrics and draws conclusions on which approach would be the best practice for a DevOps developer. The Results provide empirical evidence that offers insights into the effectiveness of each approach.

\subsection{Initial Human Efforts}
To quantify the level of human effort needed to set up a virtualization solution, we assessed several factors that affect the the infrastructure costs, such as time required to set up the infrastructure, and the time needed to start and stop the virtualization services.

We conducted tests to measure the time taken to start and stop containers and virtual machines, with four DevOps developers as our system administrators. Docker containers utilize a layered file system for image creation, resulting in a quick start time of 40 ms and stop time of 28 ms on average to create and push a basic Docker image for a container. Virtual machines, on the other hand, operate as standalone systems and take an average of 55 seconds to start and 30 seconds to stop. This is because every time a new virtual machine is created, the entire operating system must be installed from scratch. Our comparison test assumed that the operating system had already been installed inside the virtual machine and that the Docker image with Apache installed was readily available. The results of our test are presented in Table \ref{tab:human-effort-1}, Table \ref{tab:human-effort-2}, Table \ref{tab:human-effort-3}, and Table \ref{tab:human-effort-4} which clearly show that Docker containers start and stop much faster than virtual machines. As Docker containers work as processes inside the host operating system, invoking a system call to spawn a Docker container takes only milliseconds. On the other hand, each virtual machine must go through the entire generic operating system boot process, such as Power On Self Test (POST), loading GRUB, loading \texttt{initramfs}, and then starting the \texttt{init} process of the virtual machine, which takes tens of seconds. Based on our results, we concluded that Docker containers are the better choice for DevOps developers, as they save a significant amount of human effort in terms of the time required to start or stop the systems. \\
\begin{table}[h]
\centering
\caption{SysAdmin 1 Human Effort}
\label{tab:human-effort-1}
\begin{tabular}{|c|c|c|}
    \hline
    Type  & Time to start & Time to stop \\
    \hline
    Docker & 44ms & 28ms \\
    \hline
    VM & 59s & 33s \\
    \hline
\end{tabular}
\end{table}

\begin{table}[h]
\centering
\caption{SysAdmin 2 Human Effort}
\label{tab:human-effort-2}
\begin{tabular}{|c|c|c|}
    \hline
    Type  & Time to start & Time to stop \\
    \hline
    Docker & 39ms & 27ms \\
    \hline
    VM & 51s & 29s \\
    \hline
\end{tabular}
\end{table}

\begin{table}[h]
\centering
\caption{SysAdmin 3 Human Effort}
\label{tab:human-effort-3}
\begin{tabular}{|c|c|c|}
    \hline
    Type  & Time to start & Time to stop \\
    \hline
    Docker & 41ms & 31ms \\
    \hline
    VM & 54s & 28s \\
    \hline
\end{tabular}
\end{table}

\begin{table}[h]
\centering
\caption{SysAdmin 4 Human Effort}
\label{tab:human-effort-4}
\begin{tabular}{|c|c|c|}
    \hline
    Type  & Time to start & Time to stop \\
    \hline
    Docker & 43ms & 31ms \\
    \hline
    VM & 49s & 30s \\
    \hline
\end{tabular}
\end{table}

\subsection{Disk I/O Performance}
This section aims to compare the disk I/O performance of containers and virtual machines. It is crucial for DevOps developers to choose an appropriate virtualization solution when deploying I/O-heavy applications. We also compare the virtualization techniques against bare metal server. To evaluate the performance, we utilized the \texttt{fio} workload generator to run \texttt{randread} and \texttt{randwrite} operations on the disk inside both Docker containers and virtual machines. As depicted in Figure 4, containers exhibit significantly lower overhead and provide comparable performance to that of a bare metal machine, making them a preferred choice for deploying I/O-heavy applications. In contrast, virtual machines suffer from poor performance due to significant resource overhead. Hence, we can conclude that containers offer adequate disk isolation with minimal resource overhead, whereas virtual machines' performance is severely affected.

\begin{figure}[H]
\centering
\begin{tikzpicture}[scale=0.7]
  \begin{axis}[
    ybar,
    ymin=0,
    ymax=450000,
    bar width=25pt,
    ylabel=Disk IOPS,
    xtick=data,
    xticklabels={Bare metal, Containers, VM},
    nodes near coords,
    nodes near coords align={vertical},
    enlarge x limits=0.5,
    ]
    \addplot coordinates {(1, 396000) (2, 387000) (3, 309000)};
    \addplot coordinates {(1, 300000) (2, 290000) (3, 253000)};
    \legend{randread,randwrite}
  \end{axis}
\end{tikzpicture}
\caption{Disk I/O Benchmarking (higher is better)}
\label{fig:iops-performance}
\end{figure}
\vspace{-10pt} 

\subsection{Network Bandwidth Utilization}
One crucial aspect of web applications is their utilization of network resources. To measure the network bandwidth and assess network tuning, we used a network benchmark tool called netperf. By running netperf inside virtual machines and containers, we compared the results with those obtained on bare-metal. From Figure \ref{fig:network-bw}, it is evident that virtual machines perform poorly in terms of networking bandwidth, while containers offer performance similar to bare-metal. Figure \ref{fig:network-latency} presents the average latency measurement for containers and virtual machines. Therefore, a DevOps developer looking to deploy a web application would be wise to choose containers over virtual machines.

\begin{figure}[H]
\centering
\begin{tikzpicture}[scale=0.7]
    \begin{axis}[
        ybar,
        ymin=75,
        ymax=100,
        bar width=25pt,
        xlabel={},
        ylabel={Network Bandwidth (Mbps)},
        axis x line=none,
        nodes near coords,
        nodes near coords align={vertical},
        every node near coord/.append style={font=\footnotesize},
        legend style={at={(0.5,-0.2)},
          anchor=north,legend columns=-1},
        symbolic x coords={Bare metal,Containers,VM},
        xtick=data,
        enlarge x limits=0.9
        ]
        \addplot[fill=gray!40] coordinates {(Bare metal,98)};
        \addplot[fill=blue!40] coordinates {(Containers,94)};
        \addplot[fill=red!40] coordinates {(VM,91)};
        
        \legend{Bare metal, Containers, VM}
    \end{axis}
\end{tikzpicture}
\caption{Ethernet Bandwidth Benchmarking (higher is better)}
\label{fig:network-bw}
\end{figure}

\begin{figure}[H]
\centering
\begin{tikzpicture}[scale=0.7]
    \begin{axis}[
        ybar,       
        ymin=0,        
        ymax=80,        
        bar width=25pt,        
        xlabel={},        
        ylabel={Network Latency (ms)},        
        axis x line=bottom,        
        axis y line=left,        
        nodes near coords,        
        nodes near coords align={vertical},        every node near coord/.append style={font=\footnotesize},        
        legend style={at={(0.5,-0.2)},          anchor=north,legend columns=-1},        symbolic x coords={TCP\_RR,UDP\_RR},        xtick=data,  
        enlarge x limits=0.6
        ]
        
        \addplot[fill=gray!40] coordinates {(TCP\_RR,38) (UDP\_RR,36)};
        \addplot[fill=blue!40] coordinates {(TCP\_RR,49) (UDP\_RR,48)};
        \addplot[fill=red!40] coordinates {(TCP\_RR,68) (UDP\_RR,62)};
        \legend{Bare metal, Containers, VM}
    \end{axis}
\end{tikzpicture}
\caption{Network Latency Benchmarking (lower is better)}
\label{fig:network-latency}
\end{figure}

\subsection{CPU Performance}
In terms of Floating Point Operations per Second (FLOPS), we compared the performance of containers and virtual machines. To evaluate CPU performance, we utilized Intel's Linpack tool \cite{bib11}, which measures CPU by performing heavy floating point operations. Our findings indicate that containers outperform virtual machines, as they utilize the native system calls of the host operating system and use native memory swapping for high-value floating point operations. In contrast, virtual machines perform poorly due to the hypervisor's translation of the system call to the hardware. Our CPU benchmark performance on the Intel Core i7 Skylake CPU demonstrated that containers outperform virtual machines once again. Therefore, containers provide sufficient CPU isolation with little to no overhead when compared to virtual machines with bare metal machines as a reference point.

$FLOPS = sockets \times \left(\frac{cores}{sockets}\right) \times \left(\frac{cycles}{second}\right) \times \frac{FLOPS}{cycle}$ 

\begin{figure}[h]
\centering
\begin{tikzpicture}[scale=0.7]
    \begin{axis}[
        ybar,
        ymin=50,
        ymax=220,
        bar width=25pt,
        xlabel={},
        ylabel={GigaFLOPS},
        axis x line=none,
        nodes near coords,
        nodes near coords align={vertical},
        every node near coord/.append style={font=\footnotesize},
        legend style={at={(0.5,-0.2)},
          anchor=north,legend columns=-1},
        symbolic x coords={Bare metal,Containers,VM},
        xtick=data,
        enlarge x limits=0.9
        ]
        \addplot[fill=gray!40] coordinates {(Bare metal,191)};
        \addplot[fill=blue!40] coordinates {(Containers,191)};
        \addplot[fill=red!40] coordinates {(VM,139)};
        \legend{Bare metal, Containers, VM}
    \end{axis}
\end{tikzpicture}
\caption{Linpack CPU Benchmarking (higher is better)}
\label{fig:cpu}
\end{figure}

\subsection{Checkpoint, Restore, and Migration}
Our research on checkpointing and restore mechanisms in virtual machines and containers revealed that virtual machine managers such as VirtualBox, VMWare, and KVM Qemu have well-built solutions that facilitate easy checkpointing and restore operations on virtual machines \cite{bib12}. However, we found that implementing checkpointing and restore mechanisms in containers is challenging, and there are few reliable tools available for this purpose. We discovered that while Docker has native support for checkpoint restore, and some third-party tools like CRIU (Checkpoint Restore in User-space) exist, there is a lack of comprehensive documentation and community support to make it easy for DevOps developers to migrate their containers, let alone support live migration of containers.

\section{Discussion and Analysis}\label{section:analysis}
As demonstrated in our experiments, containers outperform virtual machines. Despite the advantages of containers for their reduced resource utilization, there are certain limitations that need to be considered when using them.

\subsection{Container Security}
Although containers offer numerous benefits for DevOps developers to deploy their applications, there are several scenarios where it is wise to choose virtual machines over containers \cite{bib10}. 

\subsubsection{Shared Kernel}
Firstly, containers are based on a single Linux kernel. Therefore, if an attacker targets the underlying Linux kernel and manages to bring it down, all the containers running on top of that kernel will be at risk. Although containers offer sufficient isolation, it should not be taken for granted, especially if the application requires strong isolation and security. In such cases, it is recommended to run containers inside virtual machines for added security and isolation. Unlike virtual machines, the shared kernel in containers can impact security, and a single container can bring down the whole host. Thus, using virtual machines can be a safer option when strong security is a concern.

\subsubsection{Network Access}
Containers running on a host operating system typically share the same underlying hardware. As a result, in many deployment scenarios, containers have unrestricted access through network interfaces, which can create significant security risks. This vulnerability is often overlooked, and an attacker who has gained access to one container may be able to exploit unrestricted network access to take down other containers in the cluster. It is crucial for DevOps developers to address this issue by implementing proper network security measures to prevent unauthorized access and limit the damage that can be caused by an attack.

\subsubsection{Running Containers with Privilege Mode}
Containers are typically run as daemon processes, which poses a significant security risk if the container is executed with root user privileges. In this scenario, a malicious attacker can exploit the elevated privilege level of the container to gain access to the underlying host operating system, potentially leading to the compromise of all other containers running on that host. As such, it is strongly recommended to avoid running containers with root privileges. Instead, it is advised to use a non-privileged user account with limited permissions, which can help to mitigate the risk of container-based attacks.

\subsection{Securing the Containers}
Below are some of the tools and techniques that can be used to enhance the security of containers.

\subsubsection{App Armor} A Linux Security Module (LSM) that allows administrators to assign a security profile to each program running in the system. It enables administrators to restrict the capabilities of containers and prevent them from accessing sensitive resources.

\subsubsection{BlackDuck Security} A tool that is commonly used in container inventory and mapping known security vulnerabilities to image indexes. It helps administrators to identify and remediate security issues before they are deployed into production.

\subsubsection{REMnux} An open-source Linux toolkit that assists DevOps professionals in analyzing malware and reverse engineering infected applications. It can be used to detect and remove malware from containers and their underlying images.

\subsubsection{Cilium} A network security solution that can be used to secure container applications. It provides a network policy engine that can enforce fine-grained security policies between containers.

\subsubsection{Dockscan} A tool that can be used to analyze the installation process and monitor the running containers. It can be used to detect and remediate security issues in containers and their underlying images.

\subsection{Live Container Migration}
Our attempt to test the checkpoint and restore mechanism using Checkpoint and Restore in Userspace (CRIU) inside Docker was unsuccessful. We reached out to the CRIU and Docker communities to explore the possibility of performing checkpoint restore inside containers. However, as mentioned earlier, the lack of adequate documentation and support made it difficult for us to migrate efficiently. This highlights another area where virtual machines have an edge over containers, as many virtual machine vendors already offer reliable tools and support for live migration of virtual machines.

\section{Conclusion}\label{section:conclusion}
Our study delved into the concept of virtualization through virtual machines and containers, and aimed to demonstrate the advantages and limitations of using these technologies for application deployment. As highlighted in sections V and VI, our findings reveal that containers offer lower overheads while providing similar services to virtual machines. They consume fewer system and human resources and deliver better performance improvements, making them an ideal choice for applications that require just enough isolation. However, we also emphasized that highly secured and isolated applications may require the stronger isolation and security mechanisms offered by virtual machines. Thus, the choice between virtual machines and containers largely depends on the application requirements and the DevOps developer's priorities. In conclusion, our study suggests that containers are a viable alternative to virtual machines for applications that require just enough isolation and lesser resource utilization.

\section*{Acknowledgment}
The authors would like to thank Prof. Richard Martin for his suggestions and support throughout this research work.

\bibliographystyle{IEEEtran}  
\bibliography{bibliography}

\end{document}